# Every Corporation Owns Its Image: Corporate Credit Ratings via Convolutional Neural Networks

Bojing Feng
Center for Research on Intelligent Perception and Computing
National Laboratory of Pattern Recognition, Institute of Automation, Chinese Academy of Science
Beijing, China
bojing.feng@cripac.ia.ac.cn

Wenfang Xue*
Center for Research on Intelligent Perception and Computing
National Laboratory of Pattern Recognition, Institute of Automation, Chinese Academy of Science
Beijing, China
wenfang.xue@ia.ac.cn

Bindang Xue
School of Astronautics
Beihang University
Beijing, China
xuebd@buaa.edu.cn

Zeyu Liu
Tianjin Academy for Intelligent Recognition Technologies
Tianjin, China
zeyuliu@outlook.com

*Abstract*—Credit rating is an analysis of the credit risks associated with a corporation, which reflect the level of the riskiness and reliability in investing. There have emerged many studies that implement machine learning techniques to deal with corporate credit rating. However, the ability of these models is limited by enormous amounts of data from financial statement reports. In this work, we analyze the performance of traditional machine learning models in predicting corporate credit rating. For utilizing the powerful convolutional neural networks and enormous financial data, we propose a novel end-to-end method, Corporate Credit Ratings via Convolutional Neural Networks, CCR-CNN for brevity. In the proposed model, each corporation is transformed into an image. Based on this image, CNN can capture complex feature interactions of data, which are difficult to be revealed by previous machine learning models. Extensive experiments conducted on the Chinese public-listed corporate rating dataset which we build, prove that CCR-CNN outperforms the state-of-the-art methods consistently.

## I. INTRODUCTION

With the rapid growth of the amount of financial information on the Internet, credit rating becomes fundamental for helping financial institutions to know companies well so as to mitigate credit risks [1]. It is an indication of the level of the risk in investing with the corporation and represents the likelihood that the corporation pays its financial obligations on time [2]. Therefore, it is of great importance to model the profile of the corporation. Rating agencies including Standard and Poor's, Moody's and CCXI come up with these credit ratings based on analyzing various aspects of companies' financial data and non-financial data [3]. This assessment process is very expensive and complicated, which often takes months with many experts involved to analyze all kinds of variables that reflect the reliability of a corporation. One way to reduce financial cost and time consuming is to propose a predictive model based on historical financial information.

Nowadays, machine learning models have been widely used in a range of applications including financial fields due to their capability of modeling complicated features of financial data [2]. For example, the stock price has the nature of random and nonlinear due to kinds of factors such as political events, economic conditions, and trade behaviors [4]. Despite these factors, machine learning models still have higher accuracy in estimating the stock price. Besides, machine learning techniques gained lots of interest of researchers and engineers in credit rating prediction due to its highly practical value. The work [3] builds a stack of machine learning models aiming at composing a state-of-the-art credit rating system.

Although the models above achieve satisfactory results, they still have some limitations. On the one hand, these models being proper for various research fields such as Computer Vision (CV), Natural Language Processing (NLP). But they are not for credit rating particularly. The credit rating system needs a powerful model which is designed especially for the financial field. On the other hand, previous models usually treat corporate data as 1D, so they are limited by the enormous data and feature interactions. In this work, therefore, we develop a corporate-to-image conversion to extract the 2-D feature of raw data due to these problems.

In just a few years, convolutional neural networks [5] have achieved remarkable results in the field of computer vision. Various image classification models have been proposed, from AlexNet [5] to VGG [6], then from GoogleNet v1 to v4 [7], and later ResNet [8], DenseNet [9], etc. These models are constantly improved to achieve the state-of-the-art performance.

To overcome the limitations mentioned above, we propose a novel method for Corporate Credit Ratings via Convolutional Neural Networks, CCR-CNN for brevity, to explore amounts of complicated financial data.

The main contributions of this work are summarized as follows: Firstly, we build a Chinese public-listed corporate rating dataset for research. Secondly, a new method named as a corporate-to-image conversion is developed to extract the 2-D feature of raw data. Finally, an end-to-end model named Credit Ratings via Convolutional Neural Networks (CCR-CNN) is proposed. According to the extensive experiments on this real-world dataset, our proposed CCR-CNN model achieves the state-of-the-art performance compared with other traditional machine learning methods.

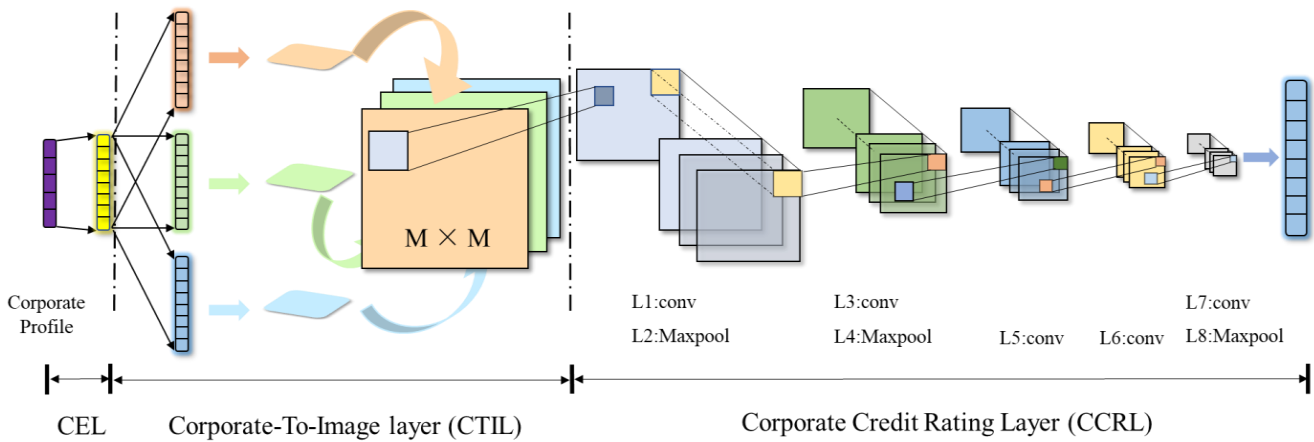

Figure 1. An Example of the architecture of our CCR-CNN model that has 5 convolution layers. CEL means the corporate embedding layer.

This paper is organized as follows. In Section II, we are going to discuss related works on the corporate credit rating. Section III will present the proposed CCR-CNN. The experiment results of CCR-CNN on real-world data will be presented in Section IV. Finally, the conclusion and future directions are introduced in Section V.

## II. RELATED WORKS

In this section, we review some related work on credit rating, including statistical models, machine learning models and hybrid models.

**Statistical Model**: Traditional statistical models including logistic regression analysis (LRA) and discriminant analysis (DA) have been come up with to build credit rating models. Huang et al. [4] replaced the linear regression function with a semi-parametric function to increase the performance. Gogas et al. [10] used an ordered probit regression model to predict bank credit rating. Recently, Pertropoulos et al. [11] proposed a model based on Student's-t Hidden Markov Models (SHMMs) to investigate the firm-related data.

**Machine Learning Model**: Due to the powerful expressive ability of machine learning models, many researchers applied machine learning techniques to predict a firm's rating. West et al. [12] built five neural network architectures to study German and Australian credit scoring data. Kim et al. [13] implemented adaptive learning networks (ALN) on both financial data and non-financial data to predict S&P credit ratings. Cao et al. [14] studied support vector machine methods on US companies from the manufacturing sector. Khandani et al. [15] studied the general classification and regression tree technique (CART) on traditional credit factors to predict consumer credit risk.

**Hybrid Model**: In addition to the statistical model and machine learning model, researchers have studied other approaches to build a credit rating model. Yeh et al. [16] combined random forest feature selection with different approaches including rough set theory (RST) and SVM. Wu & Hsu [17] proposed an enhanced decision support model combining with the relevance vector machine and decision tree. Pai et al. [18] used the Decision Tree Support Vector Machine (DTSVM) integrated TST to predict credit ratings.

## III. THE PROPOSED METHOD

In this section, we introduce the proposed CCR-CNN which applies convolution neural networks to the credit rating problems. We formulate the problem at first, then give an overview of the whole CCR-CNN, and finally describe the three layers of the model: corporate embedding layer (CEL), corporate-to-image layer (CTIL) and corporate credit rating layer (CCRL).

### A. Problem Formulation

Credit rating system aims to predict which level of credit the corporation will belong to. Here we give a formulation of this problem as below.

In credit rating, let $C = \{c_1, c_2, \cdots, c_n\}$ denotes the set consisting of all unique companies, $n$ is the number of corporates. Every corporate has a corresponding label which represents its credit level. Let $Y = \{y_1, y_2, \cdots, y_m\}$ denotes the set of the label, and $m$ represents the number of unique labels. The goal of credit rating system is to predict the corporate label according to its profile. Under the credit rating model, for the corporate $c$, we output probabilities $\hat{y}$ for all labels, where an element value of vector $\hat{y}$ is the score of the corresponding label. The model will predict the final label with the max score.

### B. Model Overview

Figure 1 illustrates the end-to-end CCR-CNN model. It is composed of three functional layers: corporate embedding layer, corporate-to-image layer and corporate credit rating layer. Firstly, every corporation is mapped into a fixed-length embedding whose length is a hyperparameter through the corporate embedding layer. Then we will get the corporate image by feeding the corporate embedding to the corporate-to image layer. Finally, corporate credit rating layer outputs the label scores by using the corporate image.

## C. Corporate Embedding Layer

For the convenience of formulation, we use $c$ to denote any corporation. It includes its financial data and non-financial data which describe the corporate profile. For numerical data, we can use it directly. However, in terms of non-numerical data, we first perform one-hot encoding, then use the embedding layer concatenating with financial data together to obtain the final corporate embedding expression $x$ via

$$x = Emb(c) \quad (1)$$

where $x \in \mathbb{R}^d$. $d$ denote the corporate embedding size.

In CCR-CNN, the choice of hidden layers has a huge impact on its performance. A straightforward solution is to use the Multi-Layer Perception (MLP) network. Despite that MLP is theoretically guaranteed to have a strong representation ability, its main drawback of having a large number of parameters can't be ignored.

## D. Corporate-to-Image Layer

In traditional data-driven methods, the data feature extraction is vital since most of the data-driven methods cannot handle the raw data [19]. However, to extract the proper features is exhausting work, and these features have great effects on the final results. In this study, an efficient feature extraction method is developed. The idea of this method is to convert the corporate embedding into images.

As shown in Figure 1, in order to obtain an $M \times M$ size image, corporate embedding $x$ is used to get three-channel embeddings $x_r, x_g, x_b$ in this conversion method.

$$x_r = W_r x + b_r \quad (2)$$
$$x_g = W_g x + b_g \quad (3)$$
$$x_b = W_b x + b_b \quad (4)$$

where $W_r, W_g, W_b \in \mathbb{R}^{M^2 \times d}$ and $b_r, b_g, b_b \in \mathbb{R}^{M^2}$ are model parameters. Then we concatenate channel embeddings into $L \in \mathbb{R}^{3 \times M^2}$ via

$$L = concat(x_r, x_g, x_b) \quad (5)$$

Finally, $P(i, j, k), i = 1, 2, 3; j = 1, \ldots, M; k = 1, \ldots, M$ denotes the pixel strength of the image, as shown in the following equation:

$$P(i,j,k) = round\left\{\frac{L(i,(j-1) \times M + k) - Min(L(i))}{Max(L(i)) - Min(L(i))} \times 255\right\}, i = 1, 2, 3 \quad (6)$$

The function $round(\bullet)$ is the rounding function and the whole pixel value has been normalized from 0 to 255. The whole image which has three channels will form a corporate image.

## E. Corporate Credit Rating Layer

Once the corporate embedding has been converted into an image, a corporate credit rating layer can be used to classify these images. This layer based on AlexNet [5] is effective and promoting, which is designed to solve the corporate image classification task.

As CNN stacks layers in a locally connected manner, this allows us to build deeper models easily than MLP. Note that due to the space limit and complicated concepts behind CNN (e.g., stride, padding, etc.), we are not ambitious to give a systematic formulation of our CCR-CNN model. Instead, without loss of generality, we briefly introduce the process of convolution on image and explain CCR-CNN of this specific setting considering its great performance in our experiments. Technically speaking, any structure of CNN and parameter setting can be employed in our CCR-CNN model.

Suppose we obtain the corporate image $P \in \mathbb{R}^{C \times M \times N}$, $C$ denotes the number of channels, $M$ denotes image size. Then we have a convolution kernel $G$ which size is $k \times k$, $G \in \mathbb{R}^{C \times k \times k}$. The convolution result $H$ can be obtained via the following formula:

$$H_{m,n} = b + \sum_{i=-\left|\frac{k}{2}\right|}^{i=-\left|\frac{k}{2}\right|} \sum_{j=-\left|\frac{k}{2}\right|}^{i=-\left|\frac{k}{2}\right|} P_{:,m+i,n+j} \bullet G_{i,j,:} \quad (7)$$

Usually, a pooling layer is added after the convolutional layer such as max pooling, mean pooling, etc., which composed a complete convolutional layer.

The following table II shows the layer configuration of the corporate credit rating layers.

TABLE I. LAYER CONFIGURATIONS OF CCRL

| Layer name | Kernel size | stride | padding |
|---|---|---|---|
| L1: Conv | 11×11×96 | 4 | 2 |
| L2:Maxpool | 3×3 | 2 | - |
| L3:Conv | 5×5×256 | 1 | 2 |
| L4:Maxpool | 3×3 | 2 | - |
| L5:Conv | 3×3×384 | 1 | 1 |
| L6:Conv | 3×3×384 | 1 | 1 |
| L7:Conv | 3×3×256 | 1 | 1 |
| L8:Maxpool | 3×3 | 2 | - |

After stacking a few convolution layers and pooling layers, we flatten the final results of L8, adding a few fully connected layers. In this way, we can get $\hat{z} \in \mathbb{R}^m$ by the final fully connected layer which has $m$ neural units according to the number of unique labels.

## F. Making Prediction and Model Training

Obtained the score $\hat{z}$ of each label, we can compute the probability for each label by applying a log softmax function to get the output vector of the model $\hat{y}$:

$$\hat{y} = \log\_\text{softmax}(\hat{z}) \quad (8)$$

where $\hat{y} \in \mathbb{R}^m$ denotes the probabilities of labels.

The loss function is defined as the cross-entropy of the prediction and the ground truth, it can be written as follows:

$$\mathcal{L}(\hat{y}) = -\sum_{i=1}^{m} y_i \log(\hat{y}_i) + (1-y_i)\log(1-\hat{y}_i) + \lambda \|\Delta\|^2 \quad (9)$$

where $y$ denotes the one-hot encoding vector of ground truth item, $\lambda$ is parameter specific regularization hyperparameters to prevent overfitting, and the model parameters of CCR-CNN are $\Delta$.

Finally, we use the Back-Propagation algorithm to train the proposed CCR-CNN model.

## IV. EXPERIMENT

In this section, we describe the datasets at first, then aim to answer the following two questions:

**RQ1**: Does the proposed CCR-CNN achieve the state-of-the-art performance compared with existing baseline algorithms?

**RQ2**: How do different sizes of the corporate image influence the performance of CCR-CNN?

### A. Dataset

Data used for model training have been collected from the annual (end of the year) financial statements of Chinese listed companies and China Stock Market & Accounting Research Database (CSMRA) for credit rating results conducted by CCXI, China Lianhe Credit Rating (CLCR), etc. Real-world data is often noisy and incomplete. Therefore, the first step of any prediction problem to credit rating, in particular, is to clean data such that we maintain as much meaningful information as possible. After data cleaning and preprocessing, which we use min-max normalization for numerical data and one-hot encoding for category data, we get 39 features and 9 rating labels: AAA, AA, A, BBB, BB, B, CCC, CC, C. The following table II will show details.

TABLE II. THE DATA FEATURES

| index | Criterion layer | | Feature Name |
|---|---|---|---|
| 1 | Financial data | Profit Capability | Net Income |
| … | | | … |
| 6 | | Operation Capability | Inventory Turn Ratio |
| … | | | … |
| 11 | | Growth Capability | Year-on-year Asset |
| … | | | … |
| 18 | | Repayment Capability | Liability To Asset |
| … | | | … |
| 25 | | Cash Flow Capability | Ebit To Interest |
| … | | | … |
| 30 | | Dupont Identity | Dupont Return on Equity |
| … | | | … |
| … | Non-Financial Data | | … |
| 39 | | | Tax Credit Rating |

Then we use Synthetic Minority Oversampling Technique (SMOTE) [20] to conduct data augmentation due to the class-imbalance problem. The final data distribution shows in following Figure 2.

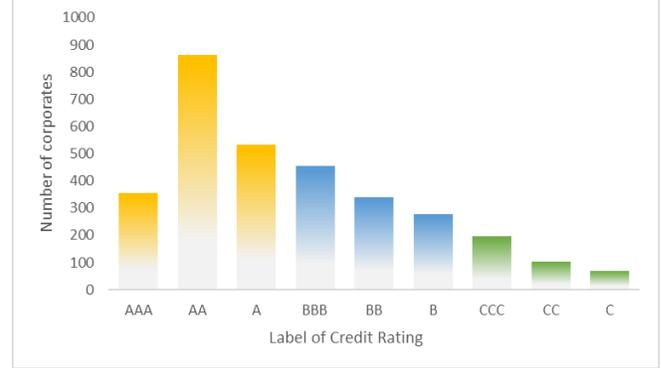

Figure 2. Data distribution after SMOTE

### B. Comparison with Baseline Methods (RQ1)

**Baseline Algorithms**

To evaluate the performance of the proposed method we compare it with the following baselines:

- **KNN:** K-Nearest Neighbor, it is one of the simplest and most effective methods in data mining classification technology.
- **LR:** Logistic regression, it is a generalized linear regression analysis model.
- **DT:** Decision tree, it is a method of approximating the value of a discrete function. It is a typical classification method.
- **RF:** Random forest, it refers to a classifier that uses multiple trees to train and predict samples [21].
- **GBDT:** Gradient Boosting Decision Tree, it is an iterative decision tree algorithm. The algorithm consists of multiple decision trees, and the conclusions of all trees are added together to make the final answer [22].
- **AdaBoost:** It is an iterative algorithm. Its core idea is to train different classifiers (weak classifiers) for the same training set, and then combine these weak classifiers to form a stronger final classifier (strong classifier) [23].
- **GaussianNB:** It is the naive Bayes whose prior distribution is Gaussian distribution [24].
- **LDA:** Linear Discriminant Analysis.
- **SVM (linear)**: Support Vector Machine with linear kernel [25].
- **SVM (rbf):** Support Vector Machine with rbf kernel [25].
- **MLP:** Multi-Layer Perceptron, A simple neural network. We use 1000 hidden units in the experiments.
- **Xgboost:** eXtreme Gradient Boosting, a scalable machine learning system for tree boosting [26].

**Evaluation Metrics**

We adopt three commonly-used metrics for evaluation, including accuracy recall and F1-score.

**Comparison with Baseline Methods**

To demonstrate the overall performance of the proposed model, we compare it with other baseline models. The overall performance in terms of accuracy recall and F1-score is shown in Table Ⅲ, with the best results highlighted in boldface.

According to the experiments, it is obvious that the proposed CCR-CNN method achieves the best performance on real dataset (Chinese public-listed corporate rating dataset). This verifies the effectiveness of the proposed method.

TABLE III. THE PERFORMACE OF CCR-CNN WITH OTHER BASELINES

| Model | Recall | Accuracy | F1-score |
|---|---|---|---|
| KNN | 0.8562 | 0.9007 | 0.8965 |
| Logistic Regression | 0.7625 | 0.8097 | 0.8195 |
| Random Forest | 0.8890 | 0.8956 | 0.9031 |
| Decision Tree | 0.8500 | 0.8271 | 0.8316 |
| GBDT | 0.9187 | 0.9264 | 0.9325 |
| AdaBoost | 0.4156 | 0.3369 | 0.3008 |
| GaussianNB | 0.5515 | 0.6554 | 0.5961 |
| LDA | 0.5875 | 0.5615 | 0.5962 |
| SVM(linear) | 0.8375 | 0.8924 | 0.8896 |
| SVM (rbf) | 0.7500 | 0.7907 | 0.8148 |
| MLP | 0.9140 | 0.9356 | 0.9296 |
| xgboost | 0.9234 | 0.9422 | 0.9413 |
| **CCR-CNN** | **0.9281** | **0.9525** | **0.9451** |

**Hyperparameter setup**

All parameters are initialized using a Xavier uniform [27] distribution with a mean of 0, and a standard deviation of 0.1. The Adam optimizer [28] is exerted to optimize these parameters, where the initial learning rate is set to 0.001 and will decay by 0.0001 after every 3 training epochs. Moreover, the L2 penalty is set to 0.00001.

*C. Find the Proper Corporate Image Size (RQ2)*

The size of corporate image in Corporate-To-image layer affects the representation ability of our CCR-CNN.

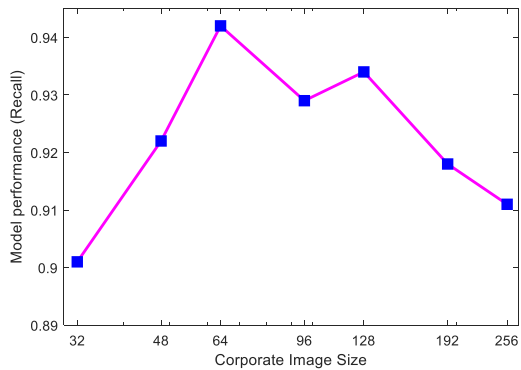

Figure 3. Model performance of different corporate image sizes.

From Figure 3 we can know that as the size of corporate image size from 32 to 64, the performance of CCR-CNN has been significantly improved. However, when we continue increasing the image size, model performance has a slight drop. What's more, CCR-CNN training time rises exponentially. Therefore, we believe it is best to set the image size to 64 according to our experiments.

As far as we know, this is the first study to build a special image for every corporation. Here we randomly choose some corporate images to show as Figure 4.

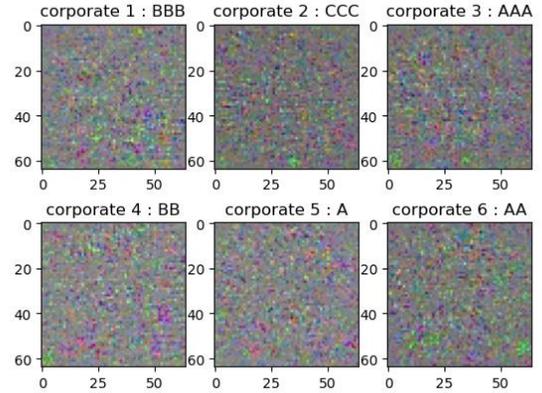

Figure 4. The corporate images with size 64

We may believe these corporate images from Figure 4 are disorganized and meaningless. However, CCR-CNN knows exactly what are they, and predict accurately their labels (credit rating results).

V. CONCLUSIONS AND FUTURE DIRECTIONS

*A. Conclusions*

In this paper, we have developed a novel corporate credit rating via the convolution neural network, CCR-CNN. We proposed a method which builds a special image for every corporation according to its enormous data. By utilizing the feature extraction power of CNN, CCR-CNN can capture and take advantage of enormous data. Besides, we have conducted a thorough empirical evaluation to investigate CCR-CNN. Extensive experiments on Chinese public-listed corporate rating dataset demonstrate the effectiveness of our model.

*B. Future Directions*

The limitations of the proposed method include the following aspects in real application. On the one hand, the results of corporate ratings may change over time. Therefore, it is necessary to build a dynamic model for corporate. Next, we will consider time factors to establish a sequential model of corporate credit ratings. On the other hand, there are connections such as cooperation between enterprises which do not exist in isolation. The connections between them will form a network. How to use such information to improve the corporate credit rating model is vital. The recent introduction of graph neural networks has brought new opportunities to solve this problem. In follow-up researches, we will build a corporate network and apply Graph Neural Networks to corporate credit rating.

ACKNOWLEDGMENT

This work is jointly supported by National Natural Science Foundation of China (62071017) and Major Projects in Tianjin Binhai New District (BHXQKJXM-PT-ZKZNSBY-2018001).

REFERENCES

[1] Golbayani P, Wang D, Florescu I. Application of Deep Neural Networks to assess corporate Credit Rating[J]. arXiv preprint arXiv:2003.02334, 2020.

[2] Golbayani P, Florescu I, Chatterjee R. A comparative study of forecasting corporate credit ratings using neural networks, support vector machines, and decision trees[J]. The North American Journal of Economics and Finance, 2020, 54: 101251.

[3] Provenzano A R, Trifiro D, Datteo A, et al. Machine Learning approach for Credit Scoring[J]. arXiv preprint arXiv:2008.01687, 2020.

[4] Huang W, Nakamori Y, Wang S Y. Forecasting stock market movement direction with support vector machine[J]. Computers & operations research, 2005, 32(10): 2513-2522.

[5] Krizhevsky A, Sutskever I, Hinton G E. Imagenet classification with deep convolutional neural networks[C]//Advances in neural information processing systems. 2012: 1097-1105.

[6] Simonyan K, Zisserman A. Very deep convolutional networks for large-scale image recognition[J]. arXiv preprint arXiv:1409.1556, 2014.

[7] Szegedy C, Liu W, Jia Y, et al. Going deeper with convolutions[C]//Proceedings of the IEEE conference on computer vision and pattern recognition. 2015: 1-9.

[8] He K, Zhang X, Ren S, et al. Deep residual learning for image recognition[C]//Proceedings of the IEEE conference on computer vision and pattern recognition. 2016: 770-778.

[9] Huang G, Liu Z, Van Der Maaten L, et al. Densely connected convolutional networks[C]//Proceedings of the IEEE conference on computer vision and pattern recognition. 2017: 4700-4708.

[10] ogas P, Papadimitriou T, Agrapetidou A. Forecasting bank credit ratings[J]. The Journal of Risk Finance, 2014.

[11] Petropoulos A, Chatzis S P, Xanthopoulos S. A Novel Corporate Credit Rating System Based on Student's-t Hidden Markov Models[J]. Expert Systems with Applications, 2016, 53(Jul.):87-105.

[12] West D. Neural network credit scoring models[J]. Computers & Operations Research, 2000, 27(11-12):1131-1152.

[13] Kim K S. Predicting bond ratings using publicly available information[J]. Expert Systems with Applications, 2005, 29(1):75-81.

[14] Cao L, Guan L K, Jingqing Z. Bond rating using support vector machine[J]. Intelligent Data Analysis, 2006, 10(3):285-296.

[15] Khandani A E, Kim A J, Lo A W. Consumer credit-risk models via machine-learning algorithms[J]. Journal of Banking & Finance, 2010, 34(11): 2767-2787.

[16] Yeh C C, Lin F, Hsu C Y. A hybrid KMV model, random forests and rough set theory approach for credit rating[J]. Knowledge-Based Systems, 2012, 33: 166-172.

[17] Wu T C, Hsu M F. Credit risk assessment and decision making by a fusion approach[J]. Knowledge-Based Systems, 2012, 35(none):102-110.

[18] Pai P F, Tan Y S, Hsu M F. Credit Rating Analysis by the Decision-Tree Support Vector Machine with Ensemble Strategies[J]. International Journal of Fuzzy Systems, 2015, 17(4):521-530.

[19] Wen L, Li X, Gao L, et al. A new convolutional neural network-based data-driven fault diagnosis method[J]. IEEE Transactions on Industrial Electronics, 2017, 65(7): 5990-5998.

[20] Chawla N V, Bowyer K W, Hall L O, et al. SMOTE: Synthetic Minority Over-sampling Technique[J]. Journal of Artificial Intelligence Research, 2002, 16(1):321-357.

[21] Breiman L. Random forests[J]. Machine learning, 2001, 45(1): 5-32.

[22] Friedman J H. Greedy Function Approximation: A Gradient Boosting Machine[J]. Annals of Statistics, 2001, 29(5):1189-1232.

[23] Yoav Freund and Robert E Schapire. 1997. A Decision-Theoretic Generalization of On-Line Learning and an Application to Boosting. J. Comput. Syst. Sci. 55, 1 (Aug. 1997), 119–139.

[24] Chan T F, Golub G H, LeVeque R J. Updating formulae and a pairwise algorithm for computing sample variances[C]//COMPSTAT 1982 5th Symposium held at Toulouse 1982. Physica, Heidelberg, 1982: 30-41.

[25] Optimizat P J S M. ion: A Fast Algorithm for Training Support Vector Machines[J]. Advances jn KerneI Metbods SuppOrt Vector I earrljng, M IT Press, Cambridge, MA, 1999: 185-208.

[26] Chen T, Guestrin C. Xgboost: A scalable tree boosting system[C]//Proceedings of the 22nd acm sigkdd international conference on knowledge discovery and data mining. 2016: 785-794.

[27] Glorot X, Bengio Y. Understanding the difficulty of training deep feedforward neural networks[C]//Proceedings of the thirteenth international conference on artificial intelligence and statistics. 2010: 249-256.

[28] Kingma D P, Ba J. Adam: A method for stochastic optimization[J]. arXiv preprint arXiv:1412.6980, 2014.